\documentclass[letterpaper,twocolumn,prl,aps,superscriptaddress,amsmath,amssymb,floatfix]{revtex4-1}
\usepackage{mathptmx}
\usepackage[latin9]{inputenc}
\usepackage{color}
\usepackage{verbatim}
\usepackage{float}
\usepackage{amsmath}
\usepackage{amssymb}
\usepackage{graphicx}
\usepackage{esint}
\usepackage[T1]{fontenc}
\usepackage[unicode=true,
 bookmarks=true,bookmarksnumbered=false,bookmarksopen=false,
 breaklinks=false,pdfborder={0 0 1},backref=false,colorlinks=true]
 {hyperref}
\hypersetup{
 linkcolor=magenta,urlcolor=blue,citecolor=blue,pdfstartview={FitH},hyperfootnotes=false}

\makeatletter

\usepackage{textcomp}
\usepackage{epstopdf}

\pdfpageheight\paperheight
\pdfpagewidth\paperwidth

\@ifundefined{textcolor}{}{%
 \definecolor{BLACK}{gray}{0}
 \definecolor{WHITE}{gray}{1}
 \definecolor{RED}{rgb}{1,0,0}
 \definecolor{GREEN}{rgb}{0,1,0}
 \definecolor{BLUE}{rgb}{0,0,1}
 \definecolor{CYAN}{cmyk}{1,0,0,0}
 \definecolor{MAGENTA}{cmyk}{0,1,0,0}
 \definecolor{YELLOW}{cmyk}{0,0,1,0}
}

\usepackage{xcolor}
\usepackage{soul}
\definecolor{blue}{rgb}{0,0,1}
\definecolor{red}{rgb}{1,0,0}
\definecolor{green}{rgb}{0,1,0}

\usepackage{soul}

\makeatother

\begin{document}
\affiliation{Institute of quantum information and technology, Nanjing University
of Posts and Telecommunications, Nanjing 210003, China}
\affiliation{CAS Key Laboratory of Quantum Information, University of Science and
Technology of China, Hefei 230026, China}
\affiliation{Broadband wireless communication and sensor network technology, key
lab of Ministry of Education, Nanjing University of Posts and Telecommunications,
Nanjing 210003, China}
\affiliation{CAS Center for Excellence in Quantum Information and Quantum Physics,
University of Science and Technology of China, Hefei 230026, China}
\affiliation{State Key Laboratory of Quantum Optics and Quantum Optics Devices,
and Institute of Opto-Electronics, Shanxi University, Taiyuan 030006,
China}

\preprint{This line only printed with preprint option}
\title{Multi-grating design for integrated single-atom trapping, manipulation
and readout}
\author{Aiping Liu }
\affiliation{Institute of quantum information and technology, Nanjing University
of Posts and Telecommunications, Nanjing 210003, China}
\affiliation{Broadband wireless communication and sensor network technology, key
lab of Ministry of Education, Nanjing University of Posts and Telecommunications,
Nanjing 210003, China}

\author{Jiawei Liu}
\affiliation{Institute of quantum information and technology, Nanjing University
of Posts and Telecommunications, Nanjing 210003, China}
\affiliation{Broadband wireless communication and sensor network technology, key
lab of Ministry of Education, Nanjing University of Posts and Telecommunications,
Nanjing 210003, China}

\author{Wei Peng}
\affiliation{Institute of quantum information and technology, Nanjing University
of Posts and Telecommunications, Nanjing 210003, China}
\affiliation{Broadband wireless communication and sensor network technology, key
lab of Ministry of Education, Nanjing University of Posts and Telecommunications,
Nanjing 210003, China}

\author{Xin-Biao Xu}
\affiliation{CAS Key Laboratory of Quantum Information, University of Science and
Technology of China, Hefei 230026, China}
\affiliation{CAS Center for Excellence in Quantum Information and Quantum Physics,
University of Science and Technology of China, Hefei 230026, China}

\author{Guang-Jie Chen}
\affiliation{CAS Key Laboratory of Quantum Information, University of Science and
Technology of China, Hefei 230026, China}
\affiliation{CAS Center for Excellence in Quantum Information and Quantum Physics,
University of Science and Technology of China, Hefei 230026, China}

\author{Xifeng Ren}
\affiliation{CAS Key Laboratory of Quantum Information, University of Science and
Technology of China, Hefei 230026, China}
\affiliation{CAS Center for Excellence in Quantum Information and Quantum Physics,
University of Science and Technology of China, Hefei 230026, China}

\author{Qin Wang}
\email{qinw@njupt.edu.cn}
\affiliation{Institute of quantum information and technology, Nanjing University
of Posts and Telecommunications, Nanjing 210003, China}
\affiliation{Broadband wireless communication and sensor network technology, key
lab of Ministry of Education, Nanjing University of Posts and Telecommunications,
Nanjing 210003, China}
\author{Chang-Ling Zou}
\email{clzou321@ustc.edu.cn}
\affiliation{CAS Key Laboratory of Quantum Information, University of Science and
Technology of China, Hefei 230026, China}
\affiliation{CAS Center for Excellence in Quantum Information and Quantum Physics,
University of Science and Technology of China, Hefei 230026, China}
\affiliation{State Key Laboratory of Quantum Optics and Quantum Optics Devices,
and Institute of Opto-Electronics, Shanxi University, Taiyuan 030006,
China}

\date{\today}

\begin{abstract}
An on-chip multi-grating device is proposed to interface single-atoms
and integrated photonic circuits, by guiding and focusing lasers to the area with $\sim$10\,$\mu$m above the chip for trapping, state manipulation and readout
of single Rubidium atoms. For the optical dipole trap, two 850\,nm laser
beams are diffracted and overlapped to form a lattice of single-atom
dipole trap, with the diameter of optical dipole trap around 2.7\,$\mu$m.
Similar gratings are designed for guiding 780\,nm probe laser to excite
and also collect the fluorescence of $^{87}$Rb atoms. Such
a device provides a compact solution for future applications of single
atoms, including the single photon source, single-atom quantum register, and sensor.
\end{abstract}
\maketitle

\section{introduction}

In the past decades, quantum information processing has been developed greatly by utilizing the quantum properties of atom and photon, where both atom internal states and photonic states can provide the carrier for encoding, storing and transporting quantum information~\cite{monroe2002quantum,Christian12,Wehner18}. A lot of effort has been dedicated to the quantum information processing based on cold atoms~\cite{Saffman2010,schafer2020tools,kaufman2021quantum}  and photons~\cite{larsen2019deterministic, zhong2020quantum}. Recently, the photonic integrated chips have been widely applied in the quantum information science and show great potentials in extending the photonic quantum information processors to tens of qubits~\cite{Liu20,bogaerts2020programmable,wang2020integrated,paraiso2021photonic}. In contrast, most previous investigations of cold atoms are implemented with conventional optical setup, which is cumbersome, sensitive to environment perturbations, and costly. Therefore, the atomic chip attracted tremendous attentions in the past decades and becomes one of the most important platforms to integrate the devices of cold atom traps and atomic states manipulation on a single chip~\cite{Hansel2001,Schumm2005,Pascal2009,Riedel2010,Machluf2013,Hattermann2017}. Similar goal is also pursued by the trapped-ion system, and the co-integration of the trapped-chip and the photonic chip was demonstrated recently~\cite{Mehta2016,Mehta2020,Niffenegger2020,Romaszko2020}. 

However, it is still challenging to integrate cold atom devices and photonic devices on a single chip. For trapping an ensemble of atoms, various methods that are compatible with the photonic chip have been proposed, such as microwire atomic potential trap~\cite{Schmiedmayer95}, grating magneto-optical trap~\cite{Liang2021} and static-magnetic trap~\cite{Reichel99}. In these platforms, the state manipulation and readout of the cold atom ensemble are still resorted to the free-space laser beam excitation and fluorescence collection. Although the enhanced photon-atom interactions have been extensively studied with photonic microstructures and single atoms recently~\cite{Kohnen2011, Stehle2011,Goban2014,kim2019,Nayak19,Beguin20,Luan2020,Lukin21S}, it is still unclear how to capture individual atoms on the photonic chip without free-space optical components. Therefore, one important challenge now is how to interface integrated photonic devices with neutral atoms tens of microns above the chip, and eventually to incorporate the single atomic trap, state manipulation and readout with on-chip photonic devices. Various atomic platforms based on diffraction gratings are proposed \cite{nshii2013surface,cotter2016design,mcgilligan2016diffraction,mcgilligan2020laser,chauhan2019photonic}, which perform the superiority for the interface between chip and the free space.

In this paper, a platform of multi-grating chip is designed for trapping, manipulating and readout single-atom. The grating transforms and focus the integrated waveguide mode to free-space light beam, thus provide an efficient approach to interface the atoms and on-chip light. By designing four gratings arranged symmetrically, single atoms could be trapped in the optical dipole trap by two 850\,nm laser beams that diffracted from two gratings, and the atoms could be excited and their emissions (780\,nm) could be collected by the other two gratings. Our numerical simulation shows that the chip-to-free-space conversion has an efficiency around $\sim 45\%$, and the laser beam has a focus width of only $\sim2.7\,\mu$m. As a result, with only 10\,mW laser power in each waveguide, an optical dipole trap depth up to 0.85\,mK for $^{87}$Rb atom could be achieved. For the atom state detection, a portion of 0.66$\%$ atomic fluorescence emission could be collected into the waveguide. Our proposal provides an experimentally feasible approach to incorporate single atoms and photonic integrated chip, which holds great potentials for future integrated atom or molecule array~\cite{kaufman2021quantum,Anderegg2019} and promises hybrid quantum optics devices~\cite{Elshaari2020,Pelucchi2021}. 

\section{structure and principle}

\begin{figure*}
\centering
\includegraphics[width=1.8\columnwidth]{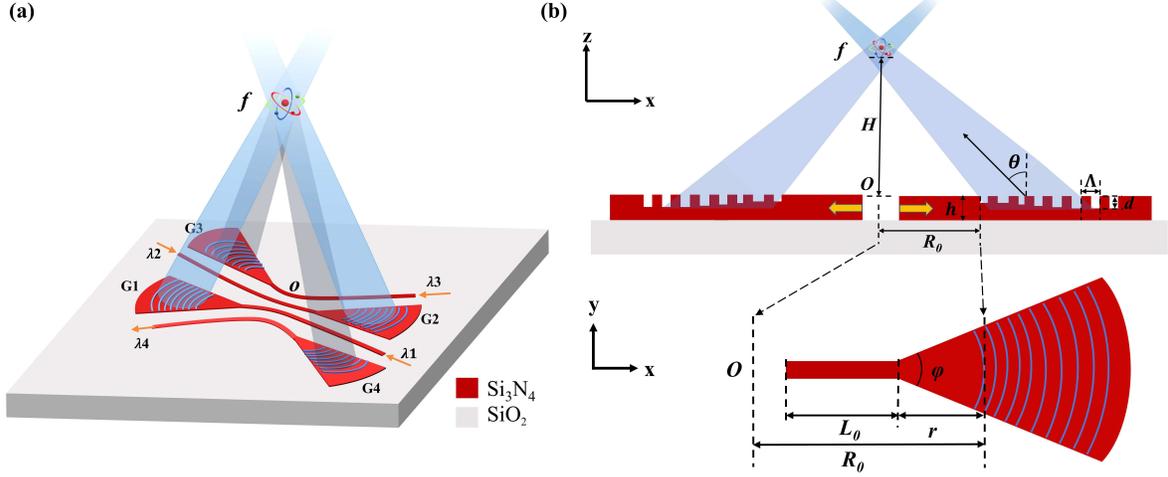}
\caption{(a) Schematic diagram of the multi-focusing-grating structure for hybrid photonic-atom chip. (b) Detailed illustration of the grating from the side view (upper panel) and the top view (bottom pannel).}
\label{fig1}
\end{figure*}

The proposed multi-grating is based on a 300\,nm thick Si\textsubscript{3}N\textsubscript{4}
wafer with SiO\textsubscript{2} as the substrate. As shown in Fig.\,\ref{fig1}(a), the designed multi-grating micro-structure is composed of four fan-shaped focusing gratings denoted by G1, G2, G3, and G4, respectively, with a common symmetric center denoted by $o$. The gratings of G1 and G2 have a rotation of $65^\circ$ relative to the gratings of G3 and G4. Due to the symmetry, it is expected that an atom at the center could be coupled with all gratings simultaneously. On one hand, the lasers guided in the waveguide could be diffracted to the free space by the gratings. As a reversal process, the photons emitted by the atom could be collected by the gratings and guided into the waveguide. In this work, we apply this hybrid photonic-atomic chip design for $^{87}$Rb: the gratings of G1 and G2 are used to diffract 850\,nm laser to form optical dipole potential well for trapping the $^{87}$Rb atom, and the gratings of G3 and G4 are used for guiding 780\,nm to excite and read out the fluorescence of the $^{87}$Rb atom. It is worth noting that the design could be extended to other atomic species by varying the geometry parameters according to the working wavelengths.

For the efficient interface between the atom and the guided photon on the chip, the conversion between the free-space beam and the waveguide mode in the waveguide is most crucial. Fig.\,\ref{fig1}(b) is the detailed geometry of the grating, by which the light propagating in the grating could be diffracted and focused. The diffraction properties of the grating are mainly determined by three parameters: period $\Lambda$, etching depth $d$ and duty cycle $\eta$ (the ratio of the unetched part to the entire cycle)~\cite{Taillaert04}. According to the first-order Bragg condition, the period $\Lambda$ of the grating is given by
\begin{equation}
\Lambda=\frac{\lambda}{n_{\mathrm{eff}}-n_{a}\mathrm{sin}\theta},
\label{eq1}
\end{equation}
where $\lambda$ is the wavelength of light in free space, $n_{\mathrm{eff}}$ is the effective refractive index of the guiding mode in the grating, $ \theta $ is the diffraction angle, and $n_{a}$ is the refractive index of the cladding above the grating with $n_{a}=1$ for air. In order to obatin a focused Gaussian-like beam, all of the diffracted light from each groove of the grating is designed to directed to the same point $f$. Then the diffraction angle $\theta_{N}$ of the $N$-th groove meets the condition
\begin{equation}
\mathrm{tan}\theta_{N}=-\frac{R_{N-1}}{H}, 
\label{eq2}
\end{equation}
where $R_{N-1}=R_{0}+\Lambda_{1}+\Lambda_{2}+\Lambda_{3}+...+\Lambda_{N-1}$ is the location of the $N$-th groove, $R_{0}$ is the distance between the symmetric center $o$ and the
first groove of the grating, and $H$ is the height of point $f$ [Fig.\,\ref{fig1}(b)]. According to Eq.\,(\ref{eq1}) and Eq.\,(\ref{eq2}), the period $\Lambda$ varies along the grating to obtain the focused beam. 

Besides, the curvature of the grating arcs should also be optimized for the focusing at $f$. The parabolic grating arc ($x\sim -y^{2}/2R$) is designed to ensure each grating arc is parallel to the phase front of wave incident on it. The intensity of the propagating light in the waveguide decreases along the grating due to the diffraction loss by the grating. An increased etching depth is designed along the grating to enhance the diffraction efficiency and thus compensate the propagating loss in the grating, since the diffraction efficiency increases with the etching depth.

\section{Results}

To validate our designs, we performed the three-dimensional finite-different time-domain (FDTD) simulations to numerically analyze the optical field distribution on the grating. In the simulation, the TE mode is adopted to be diffracted by the grating~\cite{Mak18}. We choose the width of waveguide as 550\,nm, the length and taper angle of the tapered waveguide as $r=2\,\mathrm{\mu m}$  and $\varphi=50^\circ$, respectively. The length of the grating part is 8$\,\mathrm{\mu m}$ with 18 grooves. The etching depth increases from 20\,nm to 280\,nm linearly, which can improve the diffraction efficiency of the grating \cite{wu2016fine}. Around the working wavelengths of 780\,nm and 850\,nm, the refractive index of Si\textsubscript{3}N\textsubscript{4} and SiO\textsubscript{2} are set as 1.9935 and 1.45, respectively. For simplicity, the duty cycle $\eta$ is set as 0.5.

\subsection{Single-atom trapping}

\begin{figure}
\centering\includegraphics[width=8.6cm]{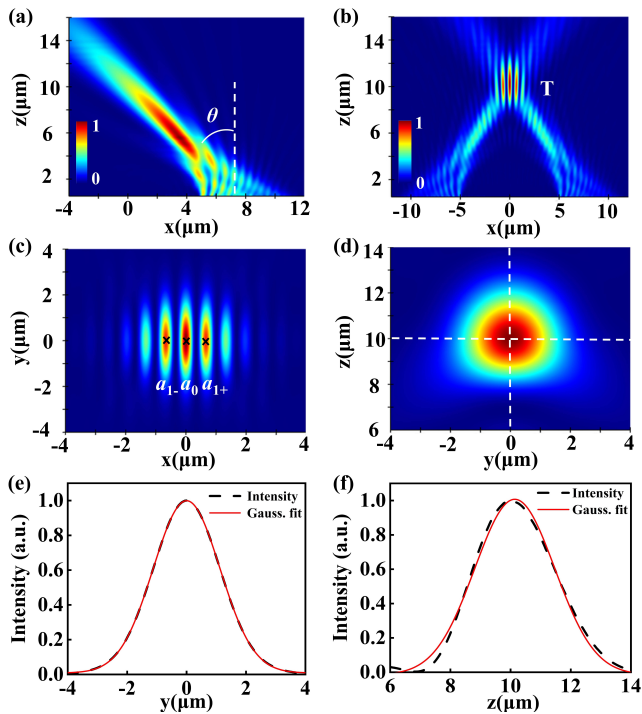}
\caption{The diffraction beams from the gratings. (a) The intensity distribution of diffraction beam from single grating in the $x-z$ plane. (b)The intensity distribution of the interference field formed by the
diffraction beams from two gratings in the $x-z$ plane, (c) in the
x-y plane, (d) in the y-z plane. The positions of trapped centers are
denoted by x in the antinodes $a_{\mathrm{1-}}$, $a_{\mathrm{0}}$, and $a_{\mathrm{1+}}$.The same colorbar for (b), (c) and (d). (e) and (f)
The intensity distributions along $y$ and $z$ directions (along dashed lines in (d)), respectively.}
\label{fig2}
\end{figure}

Figure\,\ref{fig2}(a) gives the intensity distribution of the diffraction beam from a single grating, which is designed for a focusing beam targeting at a working point height $H=10\,\mathrm{\mu m}$ with $R_{0}=4.5\,\mathrm{\mu m}$. From the intensity distribution in $x-z$ plane with $y=0$, the diffracted beam has a diffraction angle $\theta$ about -36.87$^\circ$. As expected, the diffracted beam forms a focusing Gaussian-like beam with small divergence angle,  with the waveguide-to-beam conversion efficiency of about $45.2\%$. The diffraction beam is focus on the point ($x=3.5\,\mathrm{\mu m},\, y=0,\, z=5.6\,\mathrm{\mu m}$) with a waist diameter of around $1.23\,\mathrm{\mu m}$. It should be noted that the focus of diffraction beam is deviated from the target working point($x=0,\, y=0,\, z=10\,\mathrm{\mu m}$), because there are practical limitations in designing the grating structures with a single grating arc shape. In this work, the focus point of the parabolic grating arc ($x\sim- y^{2}/2R$) is ($x=R/2,\, y=0,\, z=0$), which point is on the top of the grating and could not be accessed by other gratings. Therefore, we employing two identical gratings of G1 and G2 that are designed and arranged symmetrically with their diffraction beam incident on the same point $f$ to form a one-dimensional optical lattice at the target working point $f$ at the center. It is also guaranteed that the atoms around $f$ could be accessed by other gratings with similar design.

As shown in Fig.\,\ref{fig2}(b), two diffraction beams have the same diffraction angle of -36.87$^\circ$ and overlap with each other to form a lattice of optical dipole traps for trapping $^{87}$Rb atoms, with the antinodes (points of maximum laser intensities) locating above the center $x=0$ at a height of about $10\,\mathrm{\mu m}$. The detailed intensity distribution in the $x-y$ planes with $z=10\mu$m is shown in Figs.\,\ref{fig2}(c), where the interference of the two beams induces a fringe period of 0.67\,$\mathrm{\mu m}$ along $x$-direction. In the plane perpendicular to the lattice direction, the optical field shows a Gaussian-like beam profile [Fig.\,\ref{fig2}(d)], indicating the well-confinement to atoms in both $y$- and $z$-directions. In Figs.\,\ref{fig2}(e) and (f), the intensity distributions of dipole laser field along $y$- and $z$-directions are fitted by Gaussian functions, showing a very good agreement. The corresponding diameters of the Gaussian intensity distributions in Figs.\,\ref{fig2}(e) and (f) are 2.48\,$\mu$m and 2.66\,$\mu$m, respectively. For an input power of $850\,\mathrm{nm}$ laesr of $10\,\mathrm{mW}$, the
antinodes $a_{\mathrm{1-}}$, $a_{\mathrm{0}}$ and $a_{\mathrm{1+}}$ (as labeled in Fig.\,\ref{fig2}(c) at $x=-0.67,\,0,\,0.67\,\mathrm{\mu m}$) have electric field intensities $|E|^{2}$ of $9.3\times10^7$, $1.1\times10^8$, $9.3\times10^7$ (V/cm)\textsuperscript{2}, respectively.

For such dipole trap lattice,
%the potential When a neutral atom enters a non-uniform light field, it will experience an optical dipole force~\cite{Corwin99}
%\begin{equation}
%F=-\nabla U,
%\end{equation}
%where $U$ is optical dipole potential. 
by ignoring the Zeeman sublevels, the optical dipole potential could be estimated as~\cite{Corwin99} 
\begin{equation}
U=\frac{\hbar\gamma I_{0}}{24I_{S}}(\frac{1}{\delta_{1/2}}+\frac{2}{\delta_{3/2}}),
\end{equation}
where $\gamma/2\pi\approx6.1\,$MHz is the natural linewidth in $^{87}$Rb, $I_{S}$ is the saturation intensity, $\delta_{1/2}$ and $\delta_{3/2}$ are the detunings
between the laser frequency and the D1 and D2 transition, respectively, and $I_{0}$ is the intensity of the optical field. For the $^{87}$Rb atom with resonant wavelength at 780\,nm, the gradient force formed by optical field of 850\,nm (red-detuned with respect to both D1 and D2 transitions) is conservative force, so the $^{87}$Rb atoms could be trapped in the antinode.

\begin{figure}
\centering\includegraphics[width=8.6cm]{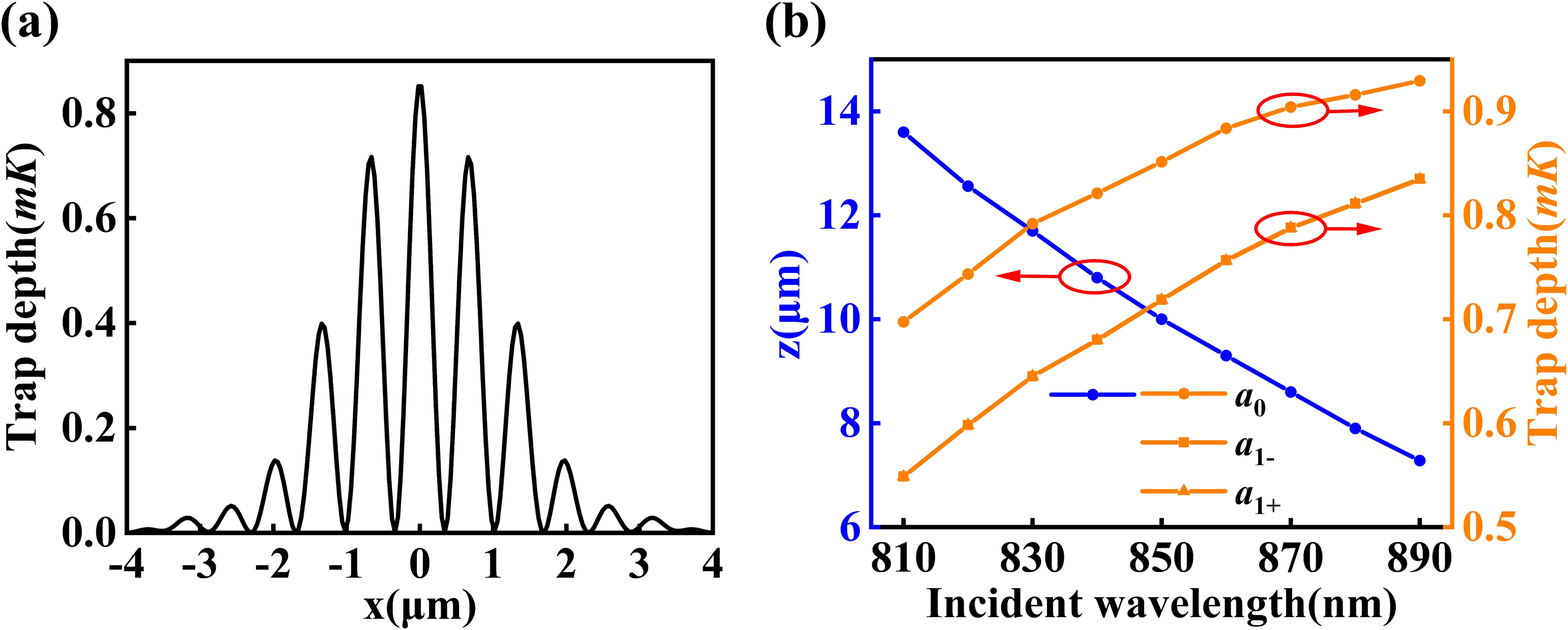}
\caption{The distribution of optical dipole potential. (a) The trap depth of optical dipole potential along x direction. (b) the
relation between the optical dipole trap and the incident wavelength.}
\label{fig3}
\end{figure}

Figure\,\ref{fig3} gives the optical dipole trap depth formed by the interference field with 10\,mW laser incident on each of the grating. The trap depth of optical dipole well along the line ($y=0, z=10\mu\textrm{m}$) for 850\,nm laer is shown in Fig.\,\ref{fig3}(a), where there are several optical dipole trap centers along $x$-direction. The trap depthes formed by the optical dipole potential in antinodes $a_{\mathrm{1-}}$, $a_{\mathrm{0}}$ and $a_{\mathrm{1+}}$ are 0.72\,mK, 0.85\,mK, and 0.72\,mK, respectively. By placing these beams into a standard MOT, where the $^{87}$Rb atoms are cooled to $~100\,\mathrm{\mu K}$, single atoms can be trapped successfully by the optical dipole well formed in these antinodes. According to previous experimental studies of single atom trapping in tweezers~\cite{Frese00}, our gratings provide dipole traps with beam waist being smaller than $3\,\mathrm{\mu m}$, thus satisfies the criteria for the atomic collision blockade~\cite{Corwin99} in the trap and we could expect at most one atom in each antinode. Since the platform is based on the transparent and optically flat photonic membrane of SiO\textsubscript{2}, the laser cooling and optical control of cold atoms can be formed above the transparent membrane~\cite{kim2019}. The platform of muti-grating is practical to trap single atom in the experiment.

Since the trap field is formed by the diffraction of gratings, the optical dipole potential is related to the incident laser wavelength. In Fig.\,\ref{fig3}(b), the relations between the optical dipole potential and the dipole laser wavelength is numerically investigated. When the wavelength varies from 810\,nm to 890\,nm, the angle and the diffraction efficiency of the beam changes. We found that the optical dipole trap of antinode $a_{\mathrm{0}}$ ($a_{\mathrm{1-}}$ or $a_{\mathrm{1+}}$) increases from 0.70(0.55)\,mK to 0.93(0.84)\,mK, and the height of the antinode $a_{\mathrm{0}}$ changes from 13.6\,$\mu$m to 7.28\,$\mu$m. Due to the structural symmetry, the positions of the antinodes are unchanged in $x$- and $y$-directions. Since the optical properties of the microstructures are determined by the dimensionless paramters $\Lambda/\lambda$ or $R_N/\lambda$, the robustness of grating against the wavelength difference also indicates the robustness of our device against the geometry parameter uncertainty due to fabrication imperfection. Therefore, our multi-grating design provides an feasible optical dipole trap with a broad bandwidth for practical experiments.

%\newpage
\subsection{Single-atom transporting}

\begin{figure}
\centering\includegraphics[width=8.6cm]{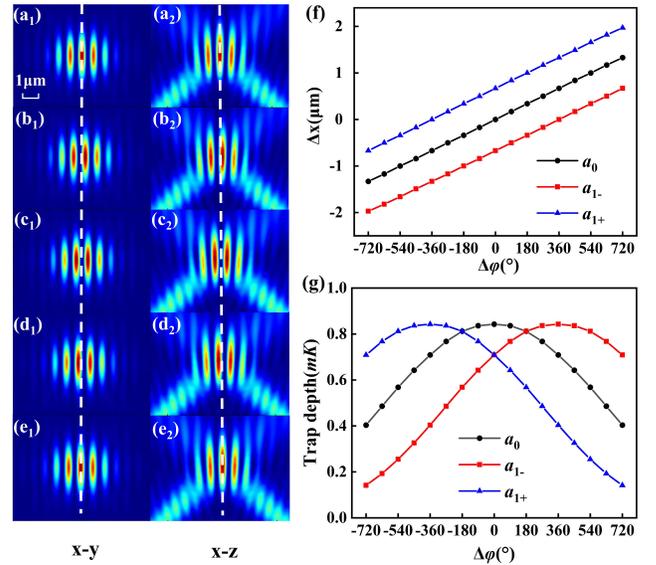}
\caption{The intensity distributions of the interference beams with different incident phase differences, (a1) - (e1) on the $x-y$ plane, (a2) - (e2) on the $x-y$ plane, with incident phase difference of 0$^\circ$, 90$^\circ$, 180$^\circ$, 270$^\circ$, and 360$^\circ$, respectively. The offsets of antinodes (f) and the optical dipole potentials (g) vary with the incident phase differences.}
\label{fig4}
\end{figure}

Since the integrated photonic chip promises the high performances thermal-optics~\cite{Idres2020} or electro-optics modulators~\cite{Liang2021}, thus it is possible to take the advantage to transport the atoms in the dipole traps. As the interference of two diffraction beams is related to the incident phase, the intensity distribution of the optical field can be manipulated by precisely controlling the laser phases. Figure\,\ref{fig4} gives the intensity distributions on the $x-y$ plane with $z=10\,\mathrm{\mu m}$ with different incident phase difference $\Delta\varphi=\varphi_{1}-\varphi_{2}$, where $\varphi_{1}$ and $\varphi_{2}$ are the incident phases of G1 and G2, respectively.
By increasing the incident phase difference $\Delta\varphi$, the
antinodes move right and their intensities are changed as shown by
Fig.\,\ref{fig4}(a1)-(e1) with incident phase difference $\Delta\varphi=0^\circ,90^\circ,180{^\circ},270{^\circ}$
and $360{^\circ}$, respectively. When the incident phase difference
$\Delta\varphi$ changes from 0 to 180$^\circ$, the antinodes are changed
to nodes, and the nodes are changed to antinodes. The interference
pattern on the x-z plane with $y=0$ has similar properties with that
on the x-y plane with $z=10\,\mu\textrm{m}$ as shown in Fig.\,\ref{fig4}(a2)-(e2).
The curves in Fig.\,\ref{fig4}(f) give the relations between the offsets
of the antinodes and the incident phase difference $ \Delta\varphi$.
When the incident phase difference increases from 0 to 360$^\circ$, the antinodes move
right linearly with a period of 0.67$\mu$m. That is the antinode
$a_{\mathrm{1-}}$ moves to the position of former antinode $a_{\mathrm{0}}$ and the antinode $a_{\mathrm{0}}$
moves to the position of former antinode $a_{\mathrm{1+}}$. The optical dipole traps formed by the antinodes also change with the incident phase
difference $\Delta\varphi$ as shown by the curves in Fig.\,\ref{fig4}(g).
The trap depths of optical dipole potential decrease with the positions of antinodes far away from the center, which is related to the incident phase difference $\Delta\varphi$. 

Benefited from that the optical dipole trap of the antinode can be
manipulated by the incident phase difference, the platform of multi-grating
can be used to trap single-atom. There is more than one antinode,
such as antinodes $a_{\mathrm{1-}}$, $a_{\mathrm{0}}$ and $a_{\mathrm{1+}}$, can trap atoms. According to the collision blockade, the loading rate of atom in a single dipole trap is around $50\%$~\cite{Schlosser02}, thus it potential to enhance the loading rate by utilizing the lattice. When there are multiple atoms loaded in the lattice, it possible to only keep a single atom at the center $a_{\mathrm{0}}$. By tuning the pase of the incident laser, the unwanted atoms could be released from the lattice as the depth of the optical dipole potential wells decreases when moving away from the center. After that, only the antinode $a_{\mathrm{0}}$ contains the trapped atoms.

\subsection{State manipulation and readout}

\begin{figure}
\centering\includegraphics[width=8.6cm]{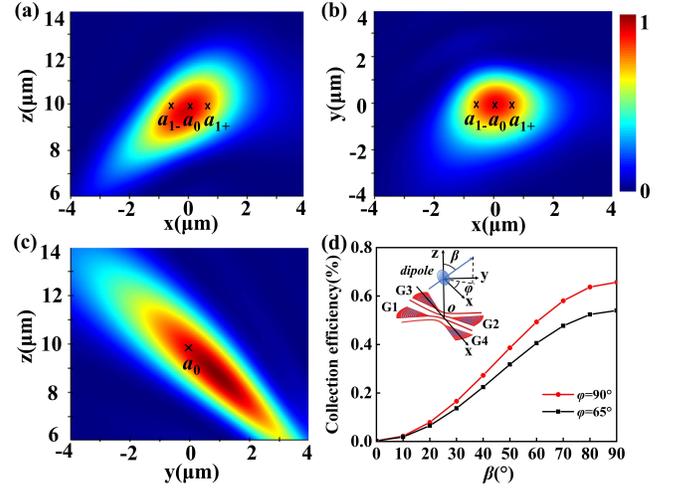}
\caption{The intensity distributions of  optical field diffracted from the grating G3  incident with 780nm laser on (a) x-z plane (b) x-y plane (c) y-z plane. (d) The collection efficiency of grating G3 varies with the angle of dipole polarization  for 780\,nm.}
\label{fig5}
\end{figure}

For an integrated photonic-atom chip, both the trapping and manipulation of atom are necessary and important for the information processing. Besides the gratings of G1 and G2 is used to trap the cold $^{87}$Rb atoms by $850\,\mathrm{nm}$ laser, similar gratings of G3 and G4 are designed for $780\,\mathrm{nm}$ light to excite and read out the fluorescence of the $^{87}$Rb atoms. Similar to the design for dipole trap laser, the G3 and G4 could be optimized with slightly different parameters. Similarly, G3 and G4 are arranged symmetrically to the center $o$, which has a rotation of 65$^\circ$ relative to the gratings of G1 and G2 as shown in Fig.\,\ref{fig1}(a). Fig.\,\ref{fig5}(a) and (b) gives the intensity distributions of the optical field on the x-y and x-z planes diffracted from the grating G3, respectively. The diffraction efficiency of grating G3 is about 47.6\%. In this multi-grating platform, the trapped $^{87}$Rb atoms can be excited by the diffraction light from the grating G3, so is the grating G4. With the incident laser power in the waveguide as low as  $1\,\mathrm{\mu W}$ at $780\,\mathrm{nm}$, the electric field strength $|E|$ on the trap centers  $a_{\mathrm{1-}}$,  $a_{\mathrm{0}}$ and $a_{\mathrm{1+}}$ are $6.7\times10^3$, $7.7\times10^3$, $7.1\times10^3$\,(V/m), and corresponding to a coherent Rabi frequency of $\Omega/2\pi=360$, $416$, and $384\,\mathrm{MHz}$, respectively. Therefore, in our configuration, the atoms could be efficiently manipulated with very low power consumption.

As a reversal process of excitation, the grating could also be utilized for collecting the atom emission. Since the waveguide mode is converted to Gaussian beam in free space, the grating is equivalent to a lens with a numerical aperture around $0.5$, while the waveguide-to-beam conversion introduces an insertion loss due to the imperfect diffraction to other directions and also to the substrates. Therefore, we also expect a considerable atom fluorescence collection efficiency from single atoms with comparing to conventional composite lens. For instances, in a typical experimental setup with a NA$=0.5$~\cite{Garthoff2021}, the collection efficiency is about $2\%$, by accounting the attenuation due to the polarization selection and the reflection of the grating, the efficiency of the fluorescence collection is estimated to be around $0.3\%$.

For a concrete estimation of the collection efficient of the single-photon emissions from the atom, we numerically solved the energy flux in the waveguide by placing a dipole at the $o$. The results for collection efficiency of grating G3 at 780\,nm are summarized in Fig.\,\ref{fig5}(d), which varies with the angle $\beta$ between dipole polarization and $z$-axis. The red and black lines correspond to the dipole polarization with $\varphi=90^\circ$ and 65$^\circ$, respectively. When the dipole polarization is parallel to the grating groove, that is $\beta=90^\circ$ and $\varphi=90^\circ$, the approximate collection efficiency of 0.66$\%$ is obtained. Considering the spontaneous emission with random polarization, the total collection efficiency is about $0.22\%$, which agrees with the above estimations. With a high collection efficiency, the designed multi-grating is feasible for the fluoresce collection, and it is potential to realize the entanglement between atoms on the same chip through the photonic Bell state measurement~\cite{Leent2021}. 

\section{conclusion}

In summary, a hybrid photonic-atomic integrated platform based on multi-grating is proposed for on-chip atom trapping, atomic states manipulation and readout. For the optical dipole trap, two 10\,mW 850\,nm laser light diffracted by two gratings to form the optical dipole trap up to 0.85\,mK for $^{87}$Rb atom. The atoms could be confined in an optical dipole trap with the potential well radius around $1.4\,\mathrm{\mu m}$, be efficiently excited by the $\mathrm{\mu W}$-level on-chip $780\,\mathrm{nm}$ laser power and be collected with efficiency of $0.66\%$. The design could be extended to other atom species, and could also be extended to even more gratings. For instance, two-dimensional atom lattice could be realized by dipole-trap lasers from three or four gratings. Further optimization of the grating structure is possible by more sophisticated chip-to-free space conversion devices, such as the metasurfaces~\cite{Huang2016}. Our device is compatible with other on-chip photonic devices, it is possible to integrate all essential optical devices for controlling and manipulating atoms on a single chip. Therefore, our work provides a unique approach to realize a scalable atom arrays, which holds great potential for the application in sensing, metrology~\cite{Norcia2019,Young2020} and quantum information processing~\cite{Saffman2010,Saffman2019,kaufman2021quantum}.

\begin{acknowledgments}
This work was supported by the Project funded by China Postdoctoral Science Foundation (SBH190004), National Key Research and Development Program of China (2018YFA0306400 and 2017YFA0304100), Leading-edge technology Program of Jiangsu Natural Science Foundation (BK20192001). X.-B.X., X.-F. R., Q. W. and C.-L.Z. are supported by the National Natural Science Foundation of China (Grants No. 11922411, 62061160487, 12074194, 12104441, and U21A6006), and the Natural Science Foundation of Anhui Provincial (Grant No. 2108085MA22), the Fundamental Research Funds for the Central Universities. The numerical calculations in this paper have been done on the supercomputing system in the Supercomputing Center of University of Science and Technology of China.
\end{acknowledgments}

\end{document}